\documentclass[runningheads,a4paper]{llncs}

\usepackage[noend]{algpseudocode}
\usepackage[ruled,linesnumbered]{algorithm2e}
\usepackage{balance}       
\usepackage{amssymb}
\usepackage{graphicx}
\usepackage{balance} 
\usepackage{subfigure} 
\usepackage{float} 
\usepackage{url}
\usepackage{ccicons}          
\usepackage[utf8]{inputenc}
\usepackage{amsmath}
\usepackage{amsfonts}
\usepackage{amssymb}
\usepackage{caption}
\usepackage{enumerate}
\usepackage{multicol}

\begin{document}

\mainmatter  

\title{User Profiling from Reviews for Accurate Time-Based Recommendations}


%
%
\author{Oznur Alkan%
\and Elizabeth Daly }
\authorrunning{User Profiling from Reviews for Accurate Time-Based Recommendations}

\institute{IBM Research, Ireland\\
\url{oalkan2@ie.ibm.com, elizabeth.daly@ie.ibm.com}}

%
%

\toctitle{Lecture Notes in Computer Science}
\maketitle

\begin{abstract}
Recommender systems are a valuable way to engage users in a system, increase participation and show them resources they may not have found otherwise. One significant challenge is that user interests may change over time and certain items have an inherently temporal aspect. As a result, a recommender system should try and take into account the time-dependant user-item relationships. However, temporal aspects of a user profile may not always be explicitly available and so we may need to infer this information from available resources. Product reviews on sites, such as Amazon, represent a valuable data source to understand why someone bought an item and potentially who the item is for. This information can then be used to  construct a dynamic user profile. In this paper, we  demonstrate utilising reviews to extract temporal information to infer the \textit{age category preference} of users, and leverage this feature to generate time-dependent recommendations. Given the predictable and yet shifting nature of age and time, we show that, recommendations generated using this dynamic aspect lead to higher accuracy compared with techniques from state of art. Mining temporally related content in reviews can enable the recommender to go beyond finding similar items or users to  potentially predict a future need of a user.  

\keywords{Recommender Systems, Time-Dependent Feature, Dynamic User Profiles}
\end{abstract}

\section{Introduction}

In many applications of recommender systems, user preferences may need to be dynamic. Preferences may change, because they themselves change, for example as a user ages, their needs shift. Alternatively, preferences may change due to their experiences, and we can potentially predict the impact of those actions. For example, if a user takes a recommended on-line learning course, the recommender algorithm can then predict that, a users skills and knowledge will have changed upon completion. There are a number of advantages in harnessing this predictive aspect in the recommender space. First, we have the ability to boot strap new items with limited information and user interaction. If a new item is suitable for a specific age group, without user interaction, we can start to recommend it to the appropriate user base. Secondly, we can leverage the predictive attributes to give more useful recommendations to users who have not been active on the system in some time. Consider job recommendations, the user may only be active every few years, however, recommendations should change for the user who's experience has increased over time, and we can exploit time-dependant features such as time spent in the current job and position.


Time-dependent user features can be seen similar to the context of a user. Research has demonstrated improved accuracy when it is taken into account by the recommender algorithm  \cite{adomavicius} through better understanding the current user dynamics regarding different settings, such as time-of day, day-of week, or  who the user is with. We consider this work complimentary to research in the context domain, where we propose to leverage the predictive aspects of the user profile, so not taking the current context but the predicted context as an input to the recommender.

Although time-dependent features may be useful inputs to a recommender system, these features may not always be explicitly available. In recent years, a variety of review-based recommender algorithms have been developed, where the goal is to extract the valuable information in user-generated textual reviews into the user modeling and recommending process \cite{reviewLit}. Reviews provide a valuable source of information describing whether items were purchased for the user themselves or for another person the item is targeted for. In addition, users can provide context information of the recipient of the item to enable others to evaluate their purchases. Therefore, online purchases represent a valuable piece of information for understanding \textit{who} the item is purchased for. In addition to this, reviews enables to keep track of the most recent preferences of the user, and this feature can lead to higher accuracy through recommending different yet more appropriate items regarding varying user needs at different time points.

In this work, we propose a natural language processing and analytic pipeline in order to infer the target age of a user, which represents the predicted age of the person, user is purchasing the items for. We store this as part of user profile as a \textit{time-dependent feature}, and recalculate it each time when a recommendation is to be generated. This dynamic aspect of the user profile enables the solution to potentially predict a future need of the user at specific time points. To realize our solution, we work on mentions of age related terms when purchasing products from Amazon.com from baby category. Given the predictable nature of age and time, we demonstrate that, this temporal information can be leveraged building upon traditional recommender algorithms to produce time-dependent age appropriate recommendations. 

The rest of the paper is organized as follows. In section 2, related state of art solutions are explored. Section 3 presents our work through detailing the main steps of the proposed framework. Evaluation of the solution is presented in section 4. The paper is concluded and possible future lines of research are discussed in section 5.

\section{Related Work}
Capturing information about users and their interests is the main function of user profiling and many works have been done in the field of recommender systems. Typically, user profiling is performed either utilizing  knowledge-based or machine learning-based methods \cite{middleton}. Knowledge-based
tecniques utilize users’ meta information, such as gender and age, to design rules for generating recommendations, whereas machine-learning-based approaches learn users’ interests automatically from their historical behaviours. Most of the state of art techniques adopt the machine-learning-based approaches for user profiling \cite{adomtuzhilin2015}. Matrix factorization (MF) based techniques is one of the most popular solutions, which learns the latent interests of a user by collaboratively factorizing the rating matrix over historically recorded user-item preferences \cite{Koren}. Most of the current MF techniques assume that users have constant interests, however, in many real-world scenarios, their interest may evolve in time. In \cite{Lu2016},  authors propose a  collaborative evolution model based probabilistic factorization of the user-item rating matrix in order to model user profiles that evolve over time; however, ratings may not always provide sufficient information to accurately model dynamic user interests.

Incorporating the valuable information in user-generated textual reviews into the user modeling and recommendation generation process can increase the accuracy of the solution at a significant extend. In \cite{1}, authors present a work in which ratings are justified using the latent rating dimensions and latent review topics. The work presented in \cite{7} combines content-based  and collaborative filtering techniques through harnessing the information of both ratings and reviews. They apply topic modeling on the reviews and align the topics with rating dimensions to improve prediction accuracy.  In \cite{8}, Factorized Latent Aspect Model (FLAME) is proposed, which combines the advantages of collaborative filtering and aspect based opinion mining. The method processes reviews so as to extract personalized preferences on different aspects and predicts users' aspect ratings on new items. In \cite{Bao2014}, a matrix factorization model, TopicMF, is presented, which uses the information semantically hidden in the review texts, and combine this with latent factors in rating data. 
Age related terms in product reviews are mined in \cite{dalyalkanAgeMatters}, in order to predict the target age range of items, where the aim was to analyse the target age of the products and how it differs from manufacturer's reported age group.

Different from existing studies presented so far, we use text-mining and prediction mechanisms to analyse dynamic aspects of user profile in order to enhance the already existing recommendation capabilities. We use the extracted temporal feature to generate better suggestions through considering users' varying interests at different time points in future. 

%


\section{The Proposed Solution}
The main motivation of the work is to mine available resources to extract a temporal aspect of user profiles, \textit{the target age}, which is not directly available but can be predicted from the user generated textual data. Specifically, we work on the reviews from Amazon.com \footnote{http://jmcauley.ucsd.edu/data/amazon/} belonging to \textit{baby} category for extracting  target age class of the users in order to understand who the user is buying items for. This is stored as a time-dependent feature and recalculated based on a later date when recommending future purchases. Although the proposed framework is applied to reviews from Amazon.com, it can be applied to any domain where user generated textual data is available and time dependent feature to be extracted is clearly defined.

The proposed solution involves four main steps: \textit{mining age related terms}, \textit{target age range prediction of items},  \textit{user target age profiling} and \textit{recommendation generation}, details of which are presented in the following subsections. 

\begin{figure}
\centering
\subfigure [Item]{
\includegraphics[width=2.2in]{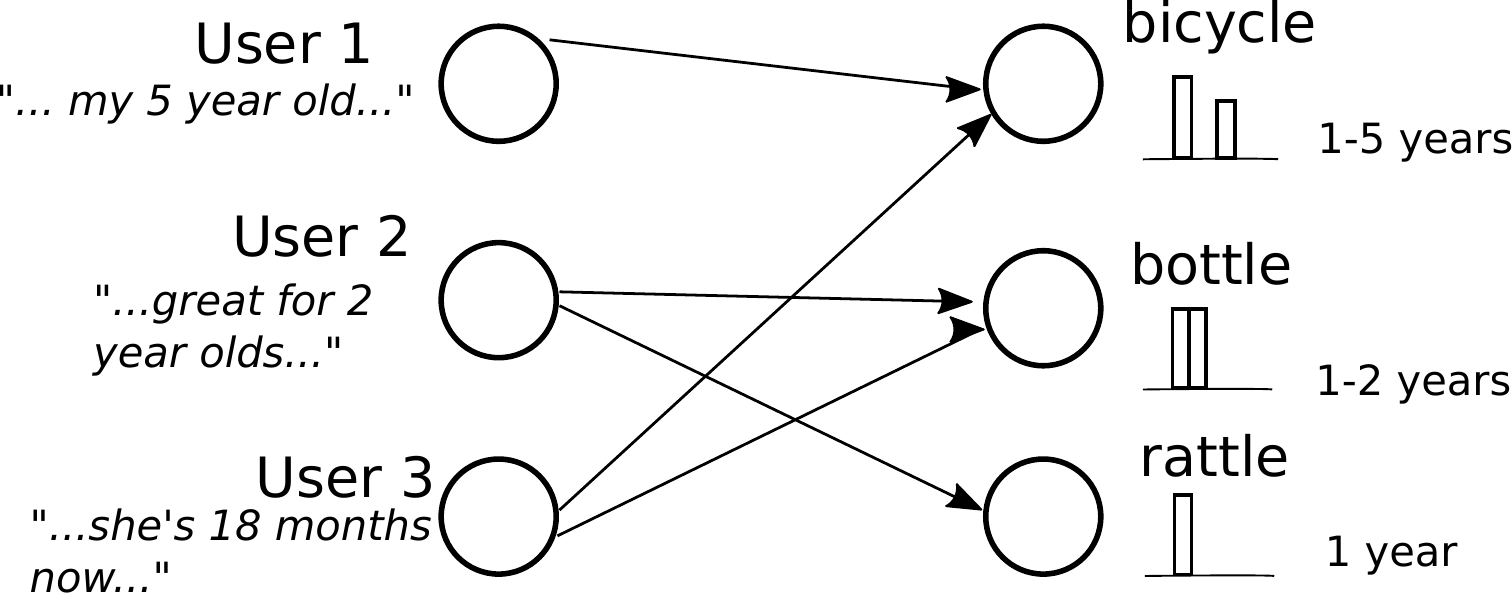} }
\hspace{1em}%
\subfigure [User] {
\includegraphics[width=2.2in]{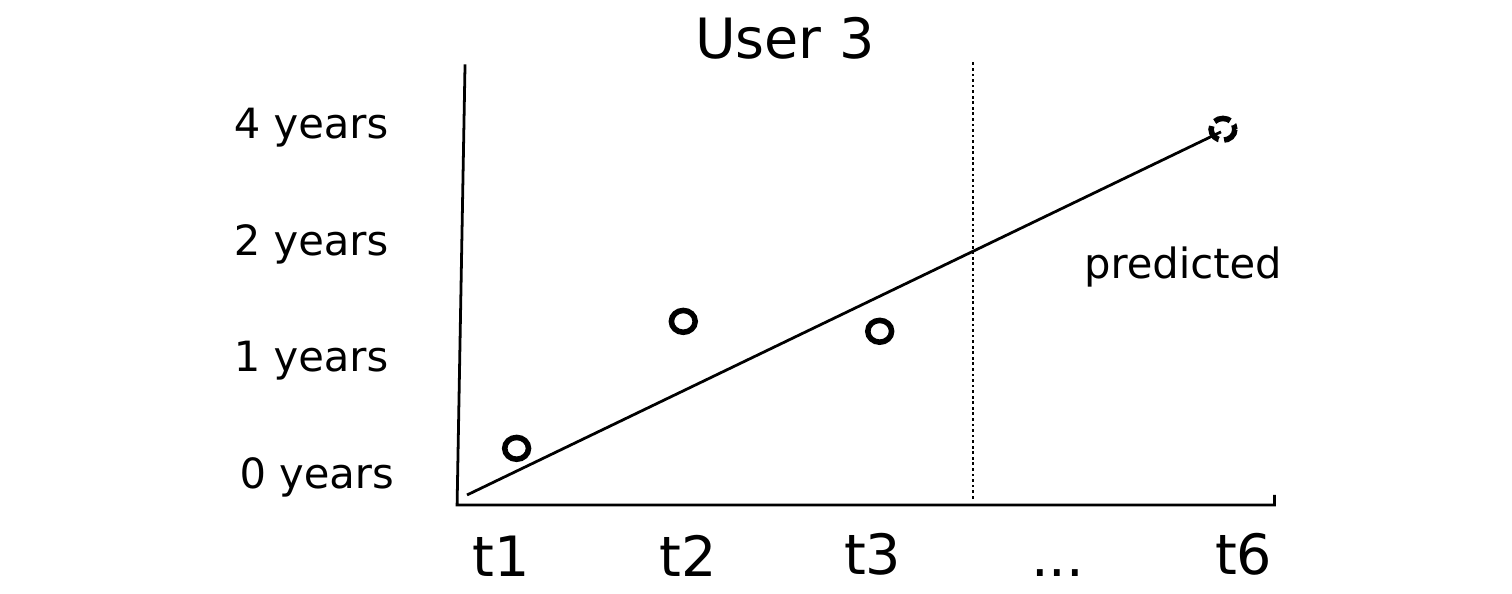} }
\caption{Age Related Term Propogation}
\label{fig:termProp}
\end{figure}

\subsection{Mining Age Related Terms}
Reviews comprise latent information about both the users and products in the sense that, they reveal how well products' features fit to users' preferences. Different from ratings, which give a general fit score, reviews have the power to give details about that fit. In this study, we investigate the age dimension of that compliance and to do so, we re-group the dataset to make analysis for both \textit{user-age} and \textit{item-age} perspectives. 

Age related terms are mined using a natural language processing pipeline in order to extract the terms needed to predict target age range of items and target age of users. Figure \ref{fig:termProp} demonstrates how age related mentions in review text are propagated to items in order to generate an age appropriateness range for products and users. In order to demonstrate this, data from Amazon.com belonging to \textit{baby} category is analysed which contains reviews, ratings and meta-data for the products. The dataset includes \textit{915,446} reviews for \textit{64,426} products from \textit{531,890} different reviewers.  

Data is preprocessed so as to extract age related terms and this process involves the following substeps that are run in sequence:
\begin{itemize}

\item \textbf{Review Selection}
Reviews are annotated and terms are extracted using the NLTK library\footnote{http://www.nltk.org/}. This step involves extracting sentences from review texts, tokenizing and POS (Part of Speech) tagging of the words. A Regular Expression tagger is built so as to extract noun phrases that are related to \textit{age} and \textit{possession}. At this step, we do not only extract age terms but we also extract possessive pronouns that exists within a close context to the age term.  The main motivation is that, possessives will enhance the prediction capability of target ages, since it reflects more clean data. 
\item \textbf{Age Related Term Selection}
After reviews are annotated and intended noun phrases are extracted, we select a subset of this extraction that mention a unit of either \textit{month, week or year}, which are our selected unit set for age feature. Since reviews are hand written and may include typos, we find possible misspelt or shorthand alternatives of selected unit set using the \textit{gensim} implementation of \textit{Word2Vec} model \footnote{https://radimrehurek.com/gensim/}. With this step,  different variations of the unit set can be found by taking the advantage of the fact that variations have similar context within a small window size. For the purposes of the evaluation of the proposed framework,  window size was set to \textit{3}. 
After this step is completed, we have a subset of the whole dataset, that mention ages, which includes \textit{739,540} reviews from \textit{447,150} reviewers for \textit{57,246} products.
\item \textbf{Term Value Normalization:}
Terms are processed, normalized, and unified at this step, which follows the following subsequent preprocessing tasks: 
\begin{itemize}
\item \textit{converting strings to integers}: ex: "three years" to "3 years", "2 and a half year" to "2.5 years"
\item \textit{unification of unit terms}: ex: "mnts" to "month", "yrs" to "years"
\item \textit{upper case to lower case conversion}
\item \textit{separating value, possessive and unit terms}: 
This step results in forming (value, unit, possessive) triples; ex: "my 3-years", "my 3years" to pair (value:3 unit:year possessive:my). For the analysis, only possessives \textit{my} and \textit{our} are considered.
\item \textit{conversion between units}: All terms are unit-converted to years. 
\end{itemize}
\end{itemize}	
\SetAlgoLined
\begin{algorithm}  \label{alg:alg1}
\KwData{\textit{RD: user-generated reviews, I: list of items}}
\KwResult{\textit{TAI: target ages of items}}

\ForAll{$item \in I$} {
$reviews \gets collect\_reviews(RD,item)$\\
$all, rt, poss, rt\_poss\gets mineAgeTerm(reviews)$ \\
$TAI[item] \gets determineTargetAgeRange(all, rt, poss, rt\_poss)$\\
}
\caption{Target Age Profiling of Items}
\end{algorithm}

\subsection{Target Age Range Prediction of Items}

Target age range of items are constructed based on the approach described in \cite{dalyalkanAgeMatters} and the outline of this process is given in Algorithm  1. The input to the process is the user generated reviews and a list of distinct items. We consider only the products that have at least \textit{4} reviews. For each item, a target age range is determined and stored in \textit{TAI} based on the following subsets of mined age terms (Line 4):
\begin{itemize}
\item \textbf{All Age Terms}: All reviews where an age related term is mentioned in the review text.
\item \textbf{Rating Based}: Only include reviews for positively rated products (reviews rated above \textit{3})
\item \textbf{Possessive Age Terms}: Only include reviews that have age related term with a possessive pronoun.
\item \textbf{Rating Based Possessive Age Terms}:  Only include reviews for positively rated products that contain age related terms with possessive pronoun
\end{itemize}
Target age of items are predicted as a range of values considering each of these four datasets separately. Phrases not related to age range of a product but has age related terms may exist in reviews which generally occur as outliers in the data. An example of such a review is "\textit{My LO is 5 months and has no problem at all using this cup himself! I used it for my girls, now 12 and 10-years-old..}" for a product of age range 0-12 months. \textit{Tukey's range test} is used to remove outliers using 5\% and 95\% as the lower and upper quartiles respectively.

\subsection {Determination of Target Age of Users}
Reviews of each user are analysed to predict the target age of them as a function of time, which is outlined in Algorithm 2. User reviews are collected (Line 2) and  processed so as to mine age related terms (Line 8). We assume that, age terms that exist together with possessive pronouns in a close context provide more accurate information regarding the target ages of users. However, not all user reviews include possessive terms, which brings a need to utilize other age related terms as well. To realize this, age related terms that exist in a close context with a possessive pronoun are collected and processed separately than the terms that exist without possessives (Lines 5-6, 10-16). For each term, both the term itself and the date of the review that contains the term are used to predict the target age, therefore, they are stored together as pairs (Lines 11,14).

Target age prediction for users involves learning the function \textit{target\_age(u, t)} for each user \textit{u} having at least \textit{k} reviews. Here the strategy followed is as follows: we sort the age terms with respect to the dates of the corresponding reviews, then we perform linear regression on the differences between each subsequent age term and each subsequent date within the data, where this preparation of the data is realized by \textit{ageTimeNormalization()} in Algorithm 2 (Lines 20,21). We need to store the date of the first review in order to use it later on during predictions at calculating time difference (Lines 18,19). Linear regression is performed utilizing possessive terms and other age related terms separately to experiment on the effect of possessives during evaluation (Lines 22,23). Since we consider dataset only for the baby category, we assume that the majority of mined age terms may reflect ages of children that the user is frequently purchasing items for. In the current work, the main motivation is to present a dynamic feature of user profile and how to utilize it by recalculating each time a recommendation is to be generated. Therefore, we simplified the current problem by predicting one target age per user. As a future work, we plan to work on multiple possible target ages extracted from user reviews for recommending items regarding different age groups.

\begin{figure}[t]
	\centering
	\includegraphics[width=8.5cm]{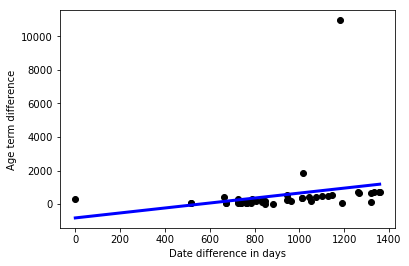}
	\caption{A sample linear regression learnt for a user having 32 possessive terms}
	\label{figure:reviewviz}
\end{figure}

As an example, in Figure \ref{figure:reviewviz}, a sample regression based analysis of possessive age terms of a user is displayed. As it can be observed from the figure, except for some outliers, the date difference is linearly increasing with the difference between the age terms mentioned at those corresponding dates of the reviews. 

\SetAlgoLined
\begin{algorithm}  
	\label{algorithm:alg2}
	\KwData{\textit{RD: user-generated reviews, \\U: list of users, \\k: term count threshold}}
	\KwResult{\textit{target\_age(u,t), $\forall u \in U$: for all users, target\_age function that returns the target age of a user for a specific time point}}
	
	\ForAll{$u \in U$} {
		$reviews \gets collect\_reviews(RD,u)$\\
		$possTerms \gets \{\}$\\	
		$terms \gets  \{\}$\\
		$possReviews \gets \{\}$\\	
		$termReviews \gets  \{\}$\\
		\ForAll{$r \in reviews$} {
			$ageTerms \gets mineAgeTerms(r)$\\
			\ForAll{$term \in ageTerms$} { 
				\eIf{term is possessive} {
					add $(reviewDate(r), term)$ pair to $possTerms$ \\
					add $r$ to $possReviews$
				}
				{
					add $(reviewDate(r), term)$ pair to $terms$ \\
					add $r$ to $termReviews$
				}
			} 
			
			$firstDate_{poss} \gets \forall r \in possReviews \{\min(reviewDate(r))\}$ \\
			$firstDate_{terms} \gets \forall r \in termReviews \{\min(reviewDate(r))\}$ \\
			$possTermsArr \gets ageTimeNormalization(firstDate_{poss}, possTerms)$\\
			$termsArr \gets ageTimeNormalization(firstDate_{terms}, terms)$\\
			$target\_age_{poss}(u,t) \gets lr(possTermsArr)$\\
			$target\_age_{term}(u,t) \gets lr(termsArr)$
		} 
	} 
	\caption{Target Age Profiling of Users}
\end{algorithm}

\SetAlgoLined
\begin{algorithm}  
	\label{algorithm:alg3}
	\KwData{\textit{u: user, t: time to generate recommendations}}
	\KwResult{\textit{recList: list of generated recommendations}}
	
	$initialList \gets cf(u)$\\
	$recList \gets \{\}$\\
	$ta\_user \gets target\_age(u,t)$\\
	\ForAll{$item \in initialList$} { 
		$(l, h) \gets targetItemAgeRange(item)$ \\
		\If{$ta\_user$ between $l$ and $h$} {
			add $item$ to $recList$ \\
		}
	} 
	\caption{Recommender Algorithm}
\end{algorithm}

\subsection{Recommendation Generation}
Recommendation generation  algorithm is outlined in Algorithm 3. For each time point \textit{t} that a recommendation is to be provided for a user \textit{u}, an initial recommendation list is constructed by utilizing a collaborative filtering technique from state of art (Line 1). Different collaborative filtering algorithms are experimented during this step, details of which are provided in section 4. The initial recommendation list is post-filtered according to the predicted target ages of the items and the result of \textit{target\_age(u,t)}, where items outside of the estimated age range for the predicted age of the target user are removed (Lines 5-8).


\begin{table}
	\centering
	\caption{Baby Review Dataset}
	\begin{tabular}{|l|c|} \hline
		SID&Sequence\\ \hline \hline
		\#reviews & $414,197$\\ \hline
		\#products & $29,230$\\ \hline
		\#users & $275,847$\\ \hline
		\#users having reviews with possessives & $771,70$\\ \hline
		avg \#reviews  per product& $9.11$\\ \hline
		avg \#review terms  per user (all terms)& $4.45$\\ \hline
		avg \#review terms with possessives per user & $1.87$\\ \hline\end{tabular}
	\label{table:baby}
\end{table}

\section{Evaluation}
\subsection{Dataset Characteristics}
Amazon reviews that exist under \textit{baby} category are used during both realization of the approach and evaluation. This category has two primary advantages. First, the purchaser is for sure buying the product for another person rather than for themselves. Second, a temporal element is present as the child increases in age. During evaluation, for the sake of accuracy, we neglect users and products having less than \textit{4} review terms. Further details of the dataset are provided in Table  \ref{table:baby}.

\subsection{Experimental setup}
We use \textit{Apache Mahout }for the evaluation along with the \textit{Rival evaluation framework} \cite{said2014rival}. For generating the initial recommendation list, three baseline recommender systems are evaluated, which are User-Based Collaborative Filtering (\textit{UB-CF}) \cite{herlocker2002empirical}, Item-Based Collaborative Filtering (\textit{IB-CF}) \cite{sarwar2001item} , and Matrix Factorization ALS-WR (\textit{MF-ALS}) \cite{zhou2008large}. In UB-CF, k-nearest neighbourhood selection with \textit{Pearson User Similarity} measure is used, whereas in \textit{IB-CF}, euclidean distance is utilized for similarity measurement. \textit{MF-ALS} utilizes matrix factorization based on alternating-least squares with weighted-$\lambda$-regularization. For all three algorithms, we set the relevance threshold to a rating of \textit{3} and use a neighbourhood size of \textit{50}. We do not evaluate all permutations of parameters, since our goal is to demonstrate the benefit of adding the temporal feature, namely, the target age to the user profile, which is reflected through post filtering the initially generated recommendation list using these baseline algorithms, rather than the performance of them.

Since the aim of the evaluation phase is to experiment on the predictive nature of age related terms, and therefore their effect as a dynamic feature of user profiles on recommendation accuracy, it is important to create the training and test set with respect to time. We leverage the \textit{Rival Temporal Splitter} which separates the data on a per user basis withholding a specified time order percentage of the user transactions. In \cite{campos2011towards}, authors demonstrated the importance of using temporal information in a realistic manner in order to evaluate recommendations. Using this scheme, \textit{80\%} of the users temporal data is used for training and the remaining \textit{20\%} used for testing.

\begin{figure}[t]
	\centering
	\includegraphics[width=1\linewidth]{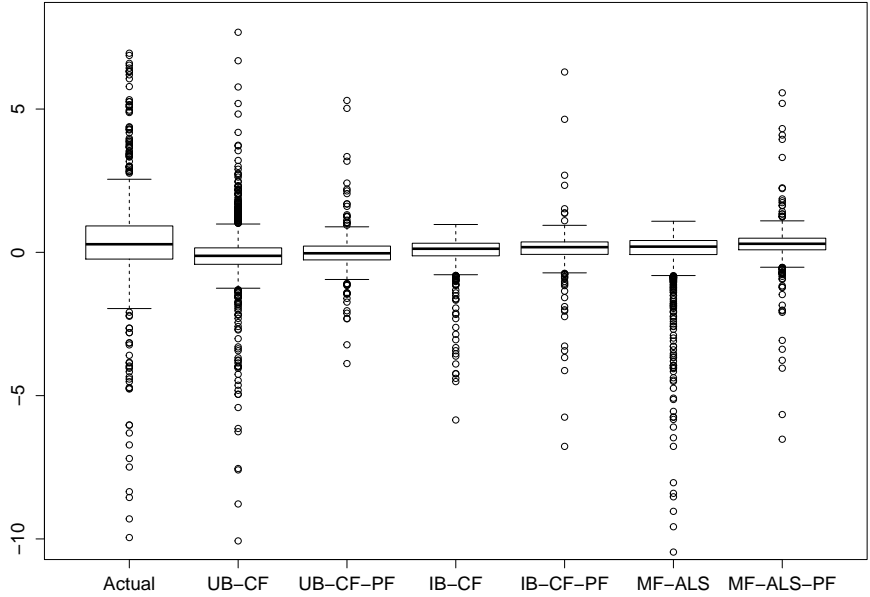}
	\caption{Temporal distribution of held out items compared to recommended items. }
	\label{figure:temporalDist}
\end{figure}

\subsection{Results and Discussion}

To understand the temporal aspect of users' purchasing behaviour, we evaluate the average age associated with items purchase in the training data set compared to the held out test dataset. Figure \ref{figure:temporalDist} shows the average estimated age of items purchased in the test set minus the average estimated age of items purchased in the training set. Here in this figure, \textit{UB-CF-PF, IB-CF-PF, MF-ALS-PF} stands for the counterparts of \textit{UB-CF, IB-CF, MF-ALS} in order, where the former utilizes target age based post filtering after generating recommendations. Considering the figure, if the product age prediction was completely accurate and users only purchased items for age ranges higher than a previous purchase, then all date points are expected to be above the negative y-axis. In reality, however, as observed from the figure, there exists a distribution that reflects some purchases in the classified age groups. The drawback of a standard collaborative filtering in terms of reflecting past purchases will be more than future purchases, which can be observed when examining results of \textit{UB-CF} and \textit{IB-CF}. Many items recommended could be classified as being purchased in the past when examining user needs and \textit{MF-ALS} also shows similar properties. However, through utilizing the temporally calculated target age of users, all three post-filtering scenarios have favoured more forward looking items. 

\begin{table}
	\centering
\caption{Example of Filtered Products based on Estimated Product Age range and Predicted User Age\label{table:age}}
	\begin{tabular}{|l|c|c|} \hline 
		Title & Estimated Age Range  & Predicted User Age\\ \hline \hline
		Fisher-Price Rainforest Friends Jumperoo	& 4 Months - 9 Months & 14 Months\\ \hline
		BRICA Fold N' Go Travel Bassinet & 2 Months - 6 Months & 14 Months\\\hline
		Similac SimplySmart Bottle & 0 Months - 12 Months & 3.5 Years\\\hline
		Fisher-Price Cruisin' Motion Soother  & 0 Months- 4 Months & 3.76 Years\\\hline
	\end{tabular}
	\label{table:wlpmResults}
\end{table}

Table \ref{table:age} shows sample recommendations that were removed through our the target age based post-filtering step. We observed examples such as removing a \textit{bassinet} recommended for a user of \textit{1.2 years old} when our predicted age relevance of the item is up to a maximum of \textit{6 months}. Similarly, \textit{a baby rocker} is removed when recommending items for a \textit{3 year old}, which is relevant considering baby category.

Table \ref{table:results} shows the results of  three baseline algorithms compared to the results from applying the post-filtering. As it can be seen from the results, metrics do support an improvement in all three algorithms by taking into account the predicted age of the user at the time, the recommendations are generated along with the estimated product age range. 

\section{Conclusions}
We have presented a framework for extracting product age related content from user reviews. These reviews are used to generate a predicting model of the target age a user, for whom the user is purchasing items for, in order to be able to recommend products that may be relevant to the user in the future. We believe this work is the first step in mining and harnessing the predictive nature of some features such as age in the realm of context based recommender systems and different than existing context based solutions, it is stored and re-evaluated at any future time point. As we have seen, collaborative filtering techniques capture current interests and relevance of the user, however if a user does not interact with the system for a time, these recommendations might become stale and no longer reflect the needs of the user. Through the propagation of age related information from products to items and items to products, we can provide users with more relevant recommended items based on the predicted current age.

\begin{table}
\centering
\caption{Evaluation Results}
\begin{tabular}{lcccc} \hline

Strategy & NDCG@10 & MAP@10 & P@10 & R@10\\  \hline \hline
UB-CF &0.0083 & 0.0048&  0.0031&0.0122\\
UB-CF-PF &0.0120 &0.0077&  0.0050&0.0156\\
IB-CF &0.0014 &0.0006& 0.0011&0.0029\\
UB-CF-PF &0.0038 &0.0026&  0.0021&0.0066\\
MF-SVD & 0.0002 & 0.0001 & 0.00007 & 0.0006\\
MF-SVD-PF & 0.0004 & 0.0002 & 0.00020 &  0.0017\\
\hline\end{tabular}
\label{table:results}
\end{table}



\balance{}



\nocite{*}
\begin{thebibliography}{4}

\bibitem {adomavicius} Adomavicius, G. and Tuzhilin, A.: Context-aware recommender systems.  In Recommender Systems Handbook, Springer US, Boston, MA, 2011, ch. 7, 217--253.

\bibitem{reviewLit} Chen, L., Chen, G., and Wang, F.: Recommender systems based on user reviews: the state of the art. User Modeling and User-Adapted Interaction 25, 2 (June 2015), 99-154.

\bibitem{proceeding3} L. Xiang and Q. Yang.: Time-Dependent Models in Collaborative Filtering Based Recommender System. 2009 IEEE/WIC/ACM International Joint Conference on Web Intelligence and Intelligent Agent Technology,450-457, 2009.	

\bibitem{url} Diao, Q., Qiu, M., Wu, C-Y., Smola, A.J., Jiang, J., Wang, C.:Jointly modeling aspects, ratings and sentiments for movie recommendation (jmars). In: 20th ACM SIGKDD international conference on Knowledge discovery and data mining, pp. 193--202, 2014.

\bibitem{dalyalkanAgeMatters} Daly, E., Alkan, O., Muller, M.: Suitable for All Ages: Using Reviews to Determine Appropriateness of Products. In: Proceedings of AAAI ICWSM, 2017.

\bibitem{herlocker2002empirical} Herlocker, J.,Konstan, J.A., Riedl, J.: An empirical analysis of design choices in neighborhood-based collaborative filtering algorithms. Information Retrieval, v.5 n.4, p.287-310, October 2002.

\bibitem{1} McAuley,J., Leskovec,J.: Hidden factors and hidden topics: understanding rating dimensions with review text. In Proceedings of the 7th ACM conference on Recommender systems, 165-172, 2013. 


\bibitem{7} Ling,G., Lyu,M.R.,King,I.: Ratings meet reviews, a combined approach to recommend. In Proceedings of the 8th ACM Conference on Recommender systems (RecSys '14). ACM, New York, NY, USA, 105-112,2014.

\bibitem{8} Wu, Y. Ester,M.:FLAME: A Probabilistic Model Combining Aspect Based Opinion Mining and Collaborative Filtering. In Proceedings of the Eighth ACM International Conference on Web Search and Data Mining (WSDM '15). ACM, New York, NY, USA, 199-208, 2015.

\bibitem{Bao2014} Bao,Y., Fang,H., Zhang,J.: TopicMF: simultaneously exploiting ratings and reviews for recommendation. In Proceedings of the Twenty-Eighth AAAI Conference on Artificial Intelligence, 2014.


\bibitem{said2014rival} Said,A., Bellogín,A.: Rival: a toolkit to foster reproducibility in recommender system evaluation. In Proceedings of the 8th ACM Conference on Recommender systems, 371-372, 2014.

\bibitem{sarwar2001item}Sarwar,B., Karypis,G., Konstan,J., Riedl,J.:  Item-based collaborative filtering recommendation algorithms. In Proceedings of the 10th international conference on World Wide Web, 285-295, 2001. 

\bibitem{zhou2008large} Zhou,Y.,Wilkinson,D., Schreiber,R., Pan,R.L: Large-Scale Parallel Collaborative Filtering for the Netflix Prize. In Proceedings of the 4th international conference on Algorithmic Aspects in Information and Management, 337-348, 2008.

\bibitem{campos2011towards} Campos,P.G., Díez,F., Sánchez-Montañés,M.:Towards a more realistic evaluation: testing the ability to predict future tastes of matrix factorization-based recommenders. In Proceedings of the fifth ACM conference on Recommender systems, 309-312, 2011.

\bibitem{middleton} Middleton, S.E., Shadbolt,N.R., Roure,D.D.D.: Ontological user profiling in
recommender systems. ACM Trans. Inf. Syst., 22(1):54–88, 2004.

\bibitem{adomtuzhilin2015} Adomavicius,G.,Tuzhilin,A.: Toward the next generation of recommender systems: A survey of the state of the art and possible extensions. IEEE Trans. Knowl. Data Eng., 17(6):734–749, 2005.

\bibitem{Koren} Koren, Y., Bell,R., Volinsky, C.: Matrix factorization techniques for recommender
systems. Computer, 42(8):30–37, 2009.

\bibitem{Lu2016} Lu,Z., Pan, S.J., Li, Y., Jiang,J., Yang, Q.: Collaborative Evolution for User Profiling
in Recommender Systems, IJCAI, New York City, USA, 2016.

\end{thebibliography}
\end{document}